# On the correlation between solar activity and large earthquakes worldwide

Vito Marchitelli[1], Paolo Harabaglia[2], Claudia Troise[3], Giuseppe De Natale[3*]

[1]Puglia Region Government, Dept. Mobility, Pub. Works, Ecology, Env., Bari, Italy

[2]Scuola di Ingegneria, Università della Basilicata, I-85100 Potenza, Italy

[3]Istituto Nazionale di Geofisica e Vulcanologia, I-80124 Naples, Italy

[*]Correspondence to: giuseppe.denatale@ingv.it

**Large earthquakes occurring worldwide have long been recognized to be non Poisson distributed, so involving some large scale correlation mechanism, which could be internal or external to the Earth. Till now, no statistically significant correlation of the global seismicity with one of the possible mechanisms has been demonstrated yet. In this paper, we analyze *20 years* of proton density and velocity data, as reported by the ISC-GEM catalogue. We found clear correlation between proton density and the occurrence of large earthquakes (*M>5.8*), with a time shift of one day. The significance of such correlation is very high, with probability to be wrong less than $10^{-5}$. The correlation increases with the magnitude threshold of the seismic catalogue. A tentative model explaining such a correlation is also proposed, in terms of the reverse piezoelectric effect induced by the applied electric field. This result opens new perspectives in seismological interpretations, as well as in earthquake forecast**.

**One Sentence Summary**: This paper puts in evidence, for the first time, a significant correlation between solar activity and earthquake occurrence.

**Introduction**

Worldwide seismicity does not follow a Poisson distribution (1), not even locally (2). Many authors have proposed statistical models to describe such a non-poissonian behavior (3-7) but none of these is really satisfactory, probably because the underlying physical process has not been really understood. Many authors have hypothesized that a tidal component might show up in earthquake activity (e.g. 8,9) but evidence has never been conclusive. Quite recently, some authors (10) suggested that earthquake occurrence might be linked to earth rotation speed variations. There is also a smaller number of researchers that studied possible links among solar activity, electro-magnetic storms and earthquakes (e.g. 11-16). The first idea that sunspots could influence the earthquake occurrence dates back 1853, and is due to the great solar astronomer Wolf (17). Since then, a number of scientists has reported some kind of relationship between solar activity and earthquake occurrence (18,16,19); or among global seismicity and geomagnetic variation (15,20) or magnetic storms (21,22). Also, some mechanisms have been proposed to justify such correlations: small changes induced by Sun-Earth coupling in the Earth's rotation speed (23); eddy electric currents induced in faults, heating them and reducing shear strength (24); or piezoelectric increase in fault stress caused by induced currents (25). However, none of these studies allowed achieving a statistically significant conclusion about the likelihood of such mechanisms. On the contrary, (26) argued that there is no convincing argument, statistically grounded, demonstrating solar-terrestrial interaction favoring earthquake occurrence. However, the large interest nowadays for possible interactions between earthquake occurrence and extra-terrestrial (mainly solar) activity, is testified for instance by the Project CSES-LIMADOU, a Chinese-Italian cooperation aimed to launch a satellite to study from space the possible influence of solar activity and ionospheric modifications on the seismicity (27). In this paper, we will try to definitively solve the problem of the existence of a correlation between solar activity and global seismicity, using a long data set and rigorous statistical analysis. Once such a correlation is demonstrated, we propose a tentative, at the moment qualitative, mechanism of possible sun-earthquakes interaction.

**Statistical assessment of Solar activity - Earthquake correlation**

Since our aim was to verify the existence of a link between solar activity and earthquakes, we considered two data sets: worldwide earthquakes, and SOHO satellite proton measurements.

As far as earthquakes are concerned, we used the ISC-GEM catalogue (28). We choose it since, at the moment, this is the only worldwide data set with homogeneous magnitude estimates that allows for sound statistical analysis. We checked its completeness for *M≥ 5.8* since *1996*. The earthquake catalogue currently (*ver. 6.0*) goes up to the end of *2015*. The earthquake catalogues we used throughout this paper, with progressively larger magnitude threshold, are reported in *Tab. 1*.

The SOHO satellite is located at the *L1* Lagrange point at about *1.5* millions of kilometers from the Earth. Hourly data in terms of proton density $\rho$ and velocity *v* are available for about *85%* of the time since early *1996*. Combining the two variables in the catalogue, we could infer, as further variables, the proton flux $\rho v$, and the dynamic pressure $\rho v^2/2$. We have therefore considered, in our analyses, four different proton variables *V*: flux, dynamic pressure, velocity, and density. We computed the average of each proton variable in consecutive daily intervals. In *Tab. 2* we report minimum, maximum, and average values for each variable *V*.

As a first step, each one of these variables *V* has been compared with the worldwide seismic events with *M≥ 5.8* in the period *1996/01/21-2015/12/31*, considering the daily number of events only. The choice of this data set is due to the fact that it is the largest one. The daily number of events is more significant than the daily total moment, since we are interested in the number of individual rupture processes, rather than in a quantity that spans several orders of magnitude. Moreover, for large numbers of events, over a few thousands, the Gutenberg Richter relation (31) is universally valid and, since earthquakes are self-similar, the number of events equivalently reflects the size of the main shock. We also chose not to decluster the event data set for two reasons. First, according to (32), it is wrong to distinguish between main events, aftershocks, and background activity; second, declustering is somewhat arbitrary but would anyway result in a completely uncorrelated catalogue, thus destroying the key information we are looking for.

Proton density and velocity vary with time, so if any correlation with earthquakes does exist, it must be found either in terms of different earthquake rates according to high/low proton values, or before/after the high or low values. We hence decided to investigate *6* conditions that are illustrated in *Tab. 3*.

*Fig. 1* shows, with an example made on 15 days of catalogue, the overall procedure and illustrates the meaning of the used conditions for the statistical tests.

Another important remark is that since we consider *4* variables, *6* conditions and, later in the discussion, *6* magnitude thresholds with different temporal windows, we choose to use non-dimensional algorithms, to facilitate comparison.

The first step consists in computing the average of $V$ ($V_{av}$). Because of the necessity of working with non-dimensional variables, we express the non-dimensional average of $V$ ($V_{av\_ad}$) as

$$V_{av\_ad} = (V_{av} - V_{min}) / (V_{max} - V_{min}) \tag{1}$$

approximated to the second significant digit. Then, we define a varying threshold, as

$$_VT = V_{min} + V_{step}(V_{max} - V_{min}) \tag{2}$$

for each variable $V$, where $V_{step}$ ranges from the average value of $V_{av\_ad}$ to *1*, with steps of *0.01*. For a given condition C, and for each $_VT$, we can count the number $D_C$ of days that satisfies the condition and the corresponding number of events $E_C$ occurring in those days. $D$ and $E$ are respectively the number of days where SOHO data are available and the total number of events that occur in those days. In this way for each $_VT$, we can simply define an event relative rate

$$R = (E_C/D_C)/(E/D). \tag{3}$$

In *Fig. 2* we show the event relative rate $R$ versus $V_{step}$, for each condition C, represented for the *4* variables: flux, dynamic pressure, velocity, and density. This approach implies that, if earthquakes do occur casually with respect to proton variables $V$, the event relative rate $R$ should oscillate around *1*, within a random uncertainty.

For most of the *CV* pairs shown in *Fig. 2*, we stopped computation at $V_{step} \approx 0.4$. This is due to the fact that, for larger threshold values, $D_C/D$ becomes smaller than *0.015*, thus giving a too poor sampling. This value has been selected so to have at least about *100* days satisfying the selected condition.

The final step consists in evaluating if *R* is significantly different from *1*, for any of the variables *V*, in any of the conditions *C* within a $_VT$ range. This means we need to devise a test starting from the assumption that earthquake occurrence is not poissonian (1-7). We choose to create $10^5$ synthetic data sets, using the real data inter-event time intervals randomly combined. This empirical approach ensures us a synthetic catalog that has exactly the same statistical properties as the actual one, since we obtain a random data set with the same survival function as the real one. The survival function gives the probability of occurrence of inter-event time intervals and is commonly used to describe the statistical properties of earthquake occurrence (e.g. 1, 4). We followed this empirical approach because, as stated above, there is no satisfactory distribution that describes inter-event time intervals in a non declustered event series. To clarify our approach, in *Fig. 3* we compare the real event survival function with a Poissonian one with identical event rate. As it is clear, the inter-arrival times of the real catalogue are markedly different from a Poisson distribution.

We wanted therefore to test whenever any random distribution could casually yield the same effects, in terms of *R* values, as the real one. Only if, for a given $_VT$, *R* is higher than any of the values $R_{rand}$ obtained by randomly distributed time intervals distributions, we consider that value as significant, thus clearly indicating correlation. This bootstrap technique corresponds to perform a statistical test with the null hypothesis that the observed correlation is only casual; given the number of $10^5$ realizations considered, we can reject the null hypothesis, for the significant cases in which no value of *R* is greater or equal to the observed one, with a probability to be wrong lower than 0.00001. In *Fig. 2* we show the statistically significant values of *R*, as formerly defined, as squares. We want to highlight this criterion is extremely rigorous (confidence level is very high, 99.999%, with respect to the normally used levels of 95%-99%), but in fact our aim is to demonstrate, beyond any reasonable doubt, if correlation between any proton variables and earthquakes does exist. For the same reason we used all the available proton data, even when a single day was preceded and followed by data lacking: this obviously led to *R* value, and hence significance, underestimation.

The analyses so far described, depicted in *Fig. 2*, show that the condition *1Dy bT* in *Tab. 2* (i.e. one day after the variable decrease below the threshold value) is the only significant one, and only for $\rho$ (density) and $\rho v$ (flux) variables. Moreover it monotonically increases as the threshold value increases, at least up to values of threshold not too high, where the sampling becomes too poor. Such an increasing trend of the R peak value is best observed for the density $\rho$, but it can be observed also, although with lower peak values, for the flux $\rho v$. We can therefore state that the most striking correlation between proton variables and global seismicity is with earthquakes occurring during the *1$^{st}$* day after the density value $\rho$ decreases below a certain threshold, in the $V_{step}$ range of *0.31-0.3*9. Such a range for $V_{step}$ corresponds to a range of proton density between 12.7 and 15.9 counts cm$^{-2}$.

As a final step, we have further checked the dependence of the observed *R* peak values on the magnitude threshold of the earthquake catalogue. We have then progressively increased the lower magnitude threshold of the used seismic catalogue according to *Tab. 1*.

*Fig. 4* clearly shows the correlation peak that becomes larger and larger with increasing magnitude cut-off. These results confirm the existence of a strongly significant correlation between worldwide earthquakes and the proton density in the ionosphere, due to solar activity.

**Discussion of statistical results**

All the obtained results point out the correlation between earthquakes and proton density is highly statistically significant, even if for catalogues with too large earthquake magnitude thresholds it does not strictly pass the significance test. This is due to the fact the three higher magnitude data sets (*M≥ 7.0,7.5,8.0*) are composed by a really small number of events (*Tab. 1*) and furthermore, for such reason, the Gutenberg-Richter relation is no longer valid.

As a final test, we wanted to check if the proton density catalogue is completely uncorrelated. We know, as stated above, that the seismic catalogue of strong earthquakes is non-Poissonian and internally correlated, so we have analyzed the proton density series to check if it were characterized by a white noise spectrum that would indicate an uncorrelated process. We simply computed the power spectrum, which is shown in Fig. 5; it is clearly very different from a white spectrum, presenting at least two sharp peaks. We performed such a computation on the longest uninterrupted time window that has a *405* days length. This evidence testifies that neither the proton density distribution is random. So, this definitively confirms that the observed correlation between the seismic catalogue and the proton density cannot be likely obtained by chance; because the likelihood that two quantities, each of them internally uncorrelated, show a clear mutual correlation only by chance is negligible. Observing the proton density power spectrum, furthermore, we note a very interesting feature: the sharpest peak is closely centred over a period around *27* days, which could be easily interpreted as the moon cycle; but, in the limit of the actual discretization of the period, it is also very close the Synodic rotation period of the sun (or Carrington rotation, (33). The second largest peak seems to be an overtone.

In conclusion, the analysis of the *1996-2015* worldwide earthquake catalogue shows a significant correlation with the measured proton density in the same period. Such correlation is described by a larger probability for earthquakes to occur during time windows *24* hours long just after a peak period (meant as a period spent over a certain threshold) in proton density due to solar activity. This kind of correlation between worldwide seismicity and solar activity has been checked also with other variables linked to solar activity, including proton velocity, dynamical pressure of protons, proton flux, and proton density. However, a significant correlation can be only observed with proton flux, besides proton density. The correlation is anyway much sharper using simple proton density, so evidencing that this is the really influent variable to determine correlation with earthquake occurrence. This correlation is shown to be statistically highly significant. The high significance of the observed correlation is also strengthened by the observation that, increasing the threshold magnitude of the earthquake catalogue, the correlation peak becomes progressively larger. The correlation between large earthquakes worldwide and proton density modulated by solar activity then appears to be strongly evident and significant.

**A possible qualitative model to explain observations**

Once a strong correlation between proton density, generated by solar wind, and large earthquakes worldwide has been assessed, the next step is to verify if a physical mechanism exists which could explain such a result. Several mechanisms have been proposed, till now, for solar-terrestrial triggering of earthquakes (see 26 for a review). Although former observations about solar-terrestrial triggering were not convincing (26), some of the formerly proposed mechanisms could explain our results, which are on the contrary statistically significant. In particular, Sobolev and Demin (34) studied the piezoelectric effects in rocks generated by large electric currents. Our observed correlation implies that a high electric potential sometimes occur between the ionosphere, positively charged by high proton density, and the Earth. Such a high potential could generate, both in a direct way or determining, by electrical induction, alterations of the normal underground potential, an electrical discharge, channeled at depth by large faults, which represent preferential, highly conductive channels. Such electric charge, passing through the fault, would generate, by reverse piezoelectric effect, a strain pulse, which, added to the fault loading and changing the total Coulomb stress, could destabilize the fault favoring its rupture. The reverse piezoelectric effect can be favored, in rocks, by the quartz minerals abundant in them. These kinds of effects, induced by high electrical potential between the ionosphere and the Earth, should likely be accompanied by electrical discharges in atmosphere, which would cause luminescence phenomena. Actually, there are numerous observations of macroscopic luminescence phenomena (named Earthquake Lights) before and accompanying large earthquakes (35). Moreover, these phenomena could also cause strong electromagnetic effects, which would be recorded as radio-waves; even such phenomena have been largely reported as accompanying, and generally preceding, large earthquakes (36). More in general, a lot of electro-magnetic anomalies, often well evident, are more and more frequently reported associated to moderate to large earthquakes (37). The recent scientific literature is full of hypotheses about how such electromagnetic effects, associated to large earthquakes, could be generated. The most debated question is if they can be considered as precursors (or maybe triggers) for large events, or they are caused by the process of slip on the faults which also generate the earthquake (38, 39). Here we suggest that the increase in the proton density in the ionosphere can qualitatively explain all these observations, and also give a physical basis to our statistical observations.

**Conclusions**

We point out this paper gives the first, strongly statistically significant, evidence for a high correlation between large worldwide earthquakes and the proton density in the ionosphere due to the solar wind. This result is extremely important for seismological research and for possible future implications on earthquake forecast. In fact, although the non-poissonian character, and hence the correlation among large scale, worldwide earthquakes was known since several decades, this could be in principle explained by several mechanisms. In this paper, we demonstrate that it can likely be due to the effect of solar wind, modulating the proton density and hence the electrical potential between the ionosphere and the Earth. Although a quantitative analysis of a particular, specific model for our observations is beyond the scope of this paper, we believe that a possible, likely physical mechanism explaining our statistical observations, is the stress/strain pulse caused by reverse piezoelectric effects. Such pulses would be generated by large electrical discharges channeled in the large faults, due to their high conductivity because of fractured and water saturated fault gauge. The widespread observations of several macroscopic electro-magnetic effects before, or however associated to large earthquakes, support our qualitative model to explain the observed, highly statistically significant, proton density-earthquakes correlation.

**Acknowledgments**

Data courtesy of the CELIAS/PM experiment on the Solar Heliospheric Observatory (SOHO) spacecraft. SOHO is a joint European Space Agency, United States National Aeronautics and Space Administration mission.

An anonymous referee is acknowledged for helping to improve the paper.

**Authors' Contributions**

P.H. elaborated statistical programs and analyses; V.M. worked the original idea; G.D.N. and C.T. wrote the core of the paper; all authors participated to the data analysis and interpretation, and contributed to write the final version.

**Additional Information**

There are no Competing Interests for any of the Authors

| *Start time* | *1996-01-21* |
|---|---|
| *End time* | *2015-12-31* |
| *SOHO Available Days* | 6472 |
| *Minimum Magnitude* | *Events occurred in SOHO Available Days* |
| *5.8* | *3922* |
| *6.0* | *2704* |
| *6.5* | *855* |
| *7.0* | *277* |
| *7.5* | *96* |
| *8.0* | *18* |

**TAB. 1.** Earthquake data sets used in this paper. Events are extracted from the ISC-GEM catalogue.

| *Variables* | *Min. daily av.* | *Av. daily av.* | *Max. daily av.* |
|---|---|---|---|
| *Prot. density $\rho$ [$cm^{-3}$]* | *0.26* | *5.69* | *40.33* |
| *Prot. velocity $v$ [$km\ s^{-1}$]* | *270* | *424* | *957* |
| *Prot. flux $\rho v$ [$cm^{-3}\ km\ s^{-1}$]* | *147* | *2314* | *16492* |
| *Prot. Dyn. Press. $\rho v^2/2$ [$cm^{-3}\ km^2\ s^{-2}$]* | *30135* | *495445* | *4965809* |

**TAB. 2.** The four variables $V$ we used in this paper. Proton density $\rho$ and velocity $v$ are from CELIAS/PM experiment on the Solar Heliospheric Observatory (SOHO). The other two are derived.

| Label | Description |
| --- | --- |
| aT | All the days with $V$ above the $_VT$ threshold. |
| bT | All the days with $V$ below the $_VT$ threshold. |
| 2lstDy aT | $2^{nd}$ to last day with $V$ above the $_VT$ threshold. |
| lstDy aT | Last day with $V$ above the $_VT$ threshold. |
| 1Dy bT | $1^{st}$ day with $V$ below the $_VT$ threshold. |
| 2Dy bT | $2^{nd}$ day with $V$ below the $_VT$ threshold. |

**TAB. 3.** The 6 conditions examined to verify an eventual correlation with earthquakes, given a threshold $_VT$ for any of the variables $V$.

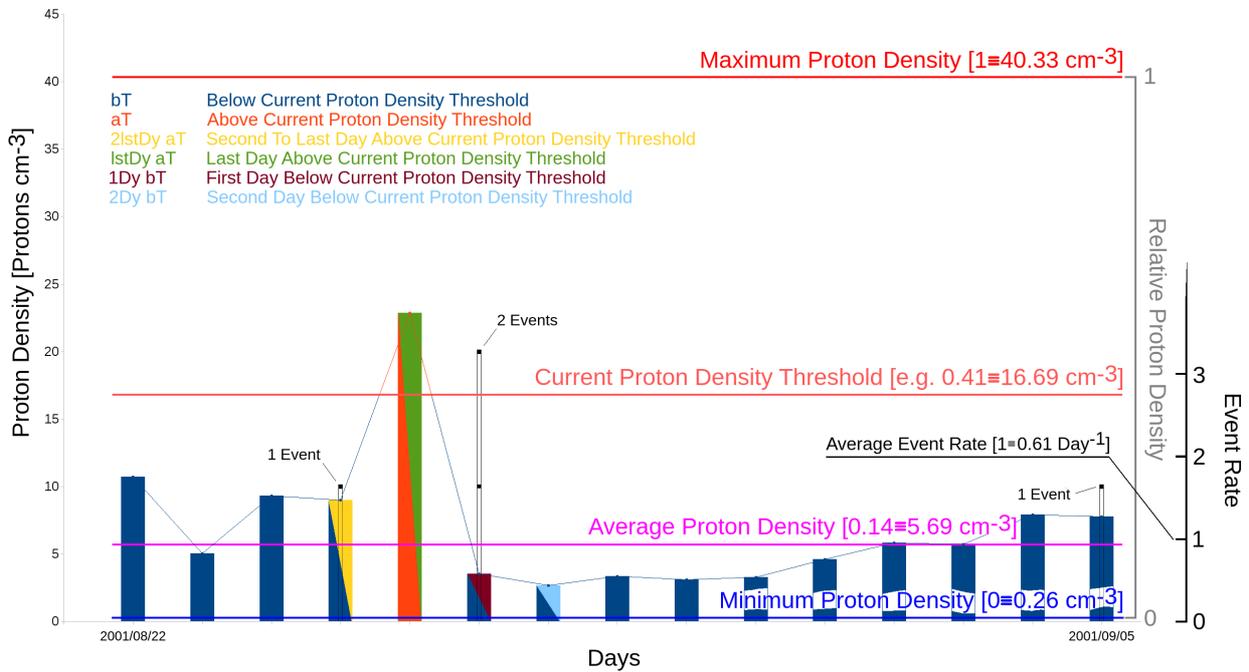

**FIG. 1**. The figure shows an example of application of the statistical method to 15 days of the catalogue. The istogram levels give the daily value of the proton density; the red line shows the level of the current density threshold (all values of it are consecutively tested). The black points indicate the occurrence of earthquakes in that day. The istogram colours indicate the conditions which are applied for the statistical tests; in particular, violet indicates the first day below the current proton density threshold (i.e. the first day after a value above the threshold), the green indicates the last day above the density threshold, and so on (as indicated in the legend). High values of earthquake frequency in one of these particular periods indicate the tendency of earthquakes to occur before, during, after (and with what time lag) a period of proton density above the current threshold. Also shown in the figure are the minimum (blue), average (purple) and maximum (intense red) values of proton density for the whole catalogue used.

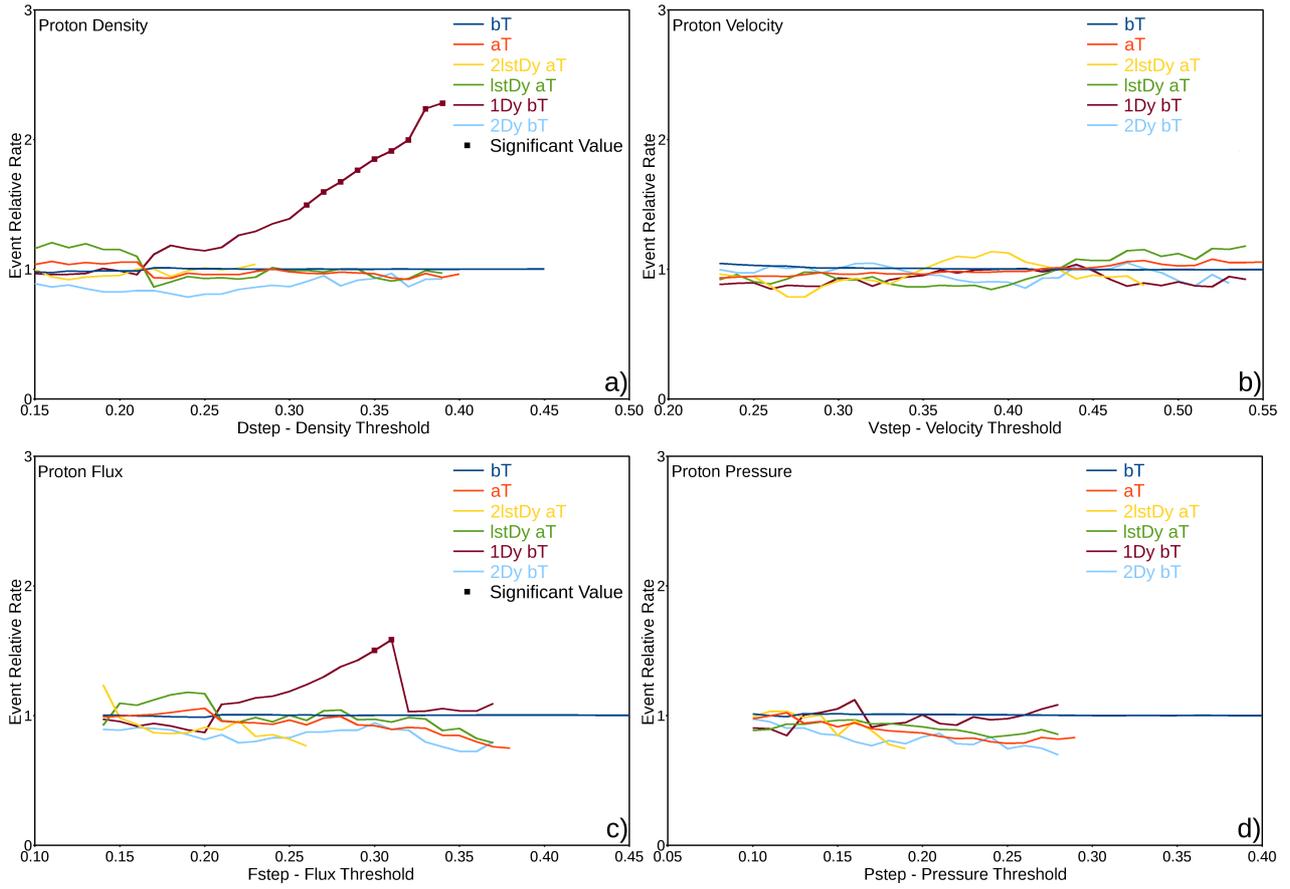

**FIG. 2**. Plots of the Event Relative Rate (see eq.2) as a function of the non-dimensional Density Threshold, for: a) proton flux; b) proton dynamic pressure; c) proton velocity; d) proton density. Different colours refer to different conditions, as explained in Tab. 2. Squares refer to values which show statistically significant correlation at a confidence level 99.999%.

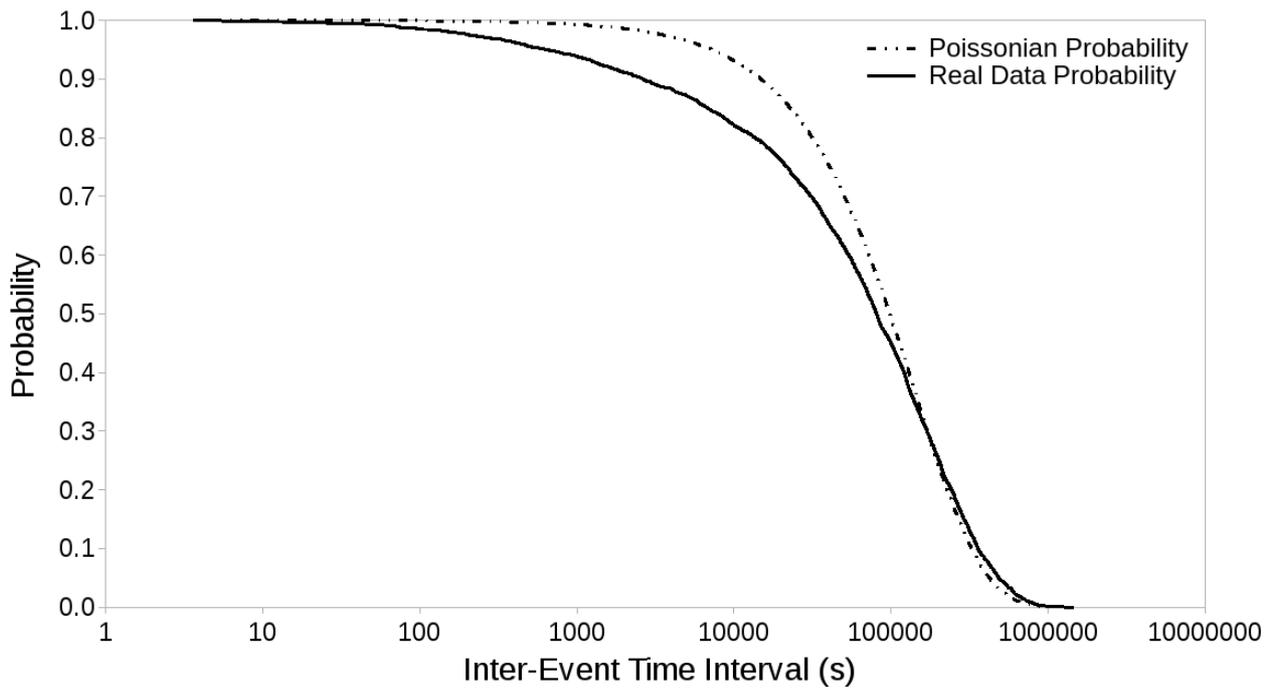

**FIG. 3**. Inter-arrival time distribution of the events in the seismic catalogue (solid line). The dotted line shows, for comparison, the expected distribution of inter-arrival times for a Poisson distribution with the same event rate.

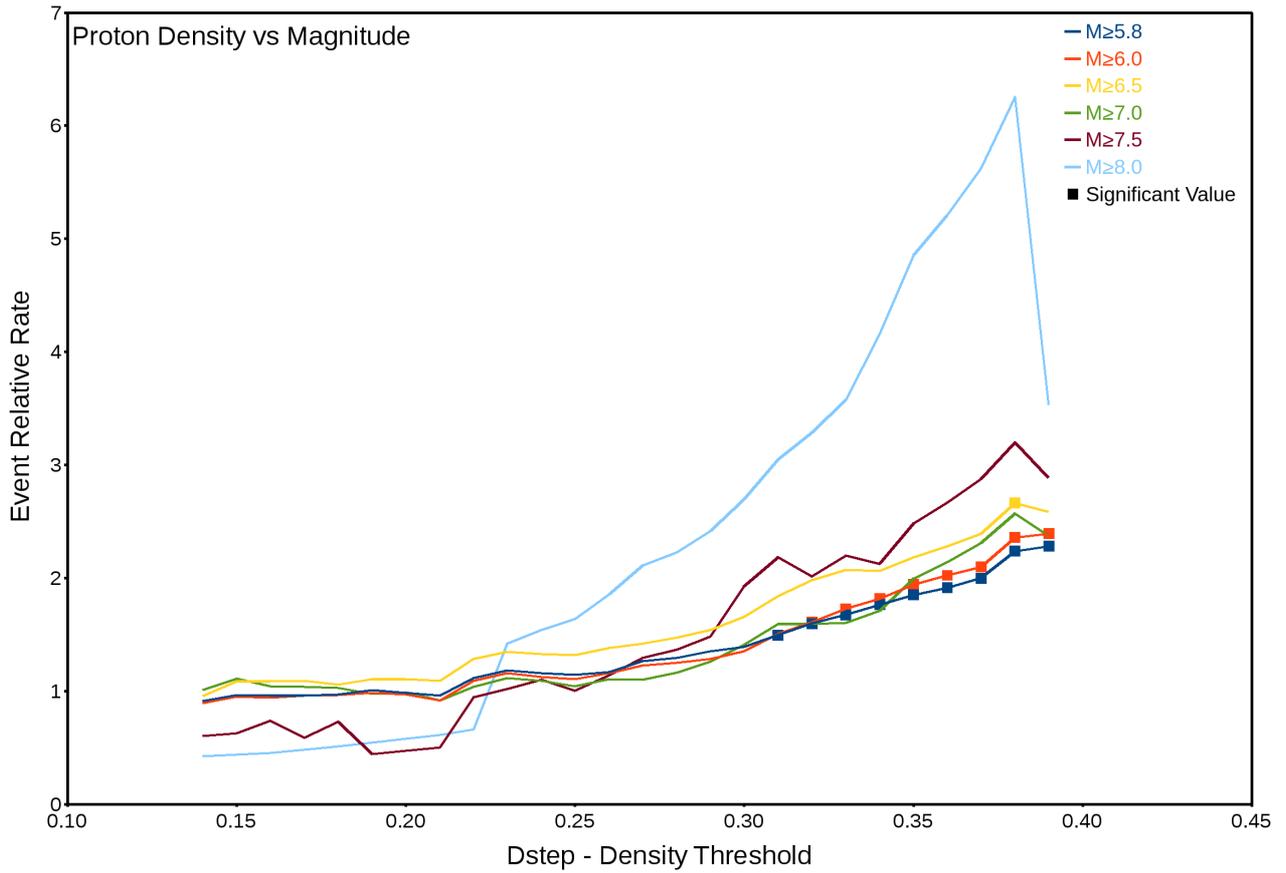

**FIG. 4.** Plots of the Event Relative Rate for proton density and the condition 1Dy bT (earthquakes occurring within 24 hours from the value of density decreasing below the threshold value). Colours indicate different lower cut-off magnitudes in the catalogue.

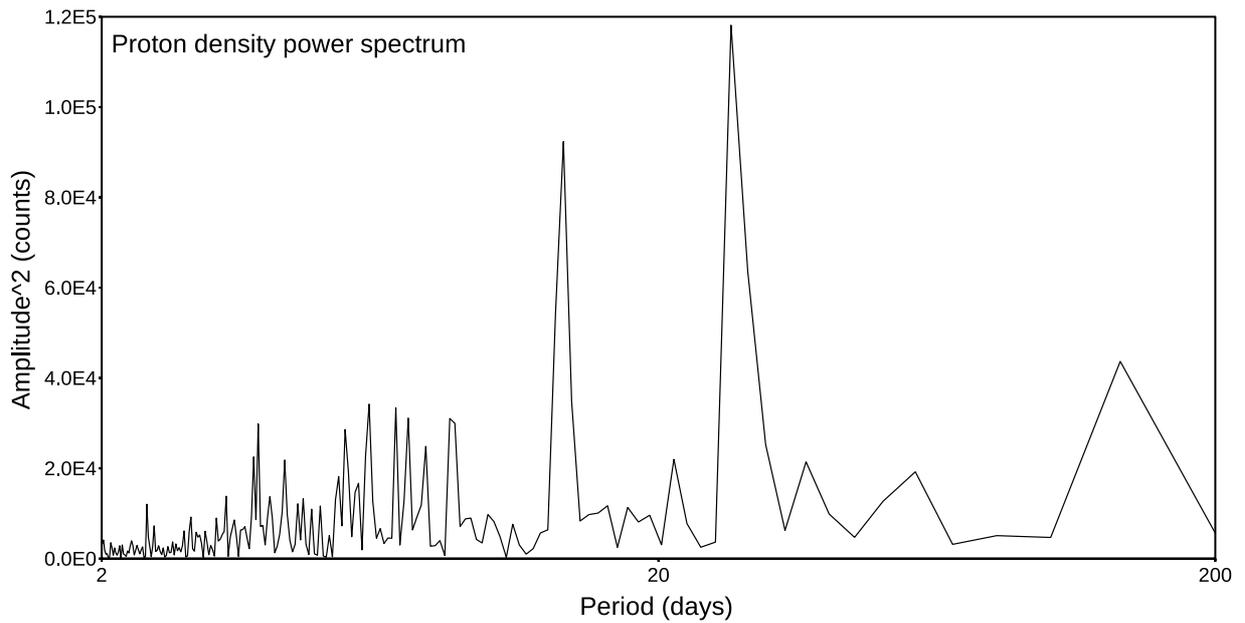

**FIG. 5**. Power spectrum of the proton density catalogue. The spectrum is computed only for the maximum consecutive period of data with no interruption, lasting 200 days.